# Configure polaritons in twisted α-MoO$_3$

M. Chen[1†], X. Lin[2†], T. Dinh[3], Z. Zheng[3], J. Shen[1], Q. Ma[3], H. Chen[2], P. Jarillo-Herrero[3], S. Dai[1]*

[1]Materials Research and Education Center, Department of Mechanical Engineering, Auburn University, Auburn, Alabama 36849, USA

[2]Interdisciplinary Center for Quantum Information, State Key Laboratory of Modern Optical Instrumentation, ZJU-Hangzhou Global Science and Technology Innovation Center, Zhejiang University, Hangzhou 310027, China

[3]Department of Physics, Massachusetts Institute of Technology, Cambridge, Massachusetts 02139, USA

[†]These authors contribute equally

*Correspondence to: sdai@auburn.edu

Moiré engineering as a configuration method to twist van der Waals materials has delivered a series of advances in electronics, magnetics and optics. Yet these advances stem from peculiar moiré superlattices which form at small specific twisting angles. Here we report the configuration of nanoscale light-matter waves – the polaritons – by twisting stacked α-phase molybdenum trioxide (α-MoO$_3$) slabs in the broad range of 0º to 90º. Our combined experimental and theoretical results reveal a variety of polariton wavefront geometry and topological transitions via the twisting. The polariton twisting configuration is attributed to the electromagnetic interaction of highly anisotropic hyperbolic polaritons in stacked α-MoO$_3$ slabs. The nano-polaritons demonstrated in twisted α-MoO$_3$ hold the promise as tailored nano-light for on-demand nanophotonic functionalities.

**Main text**

Weakly bonded atomic layers in van der Waals (vdW) systems[1-3] offer the platform to configure physical properties via twisting and stacking. This configuration has been recently investigated for the special case of moiré superlattices – the interference lattice pattern by twisting stacked vdW materials at small angles δ – to configure electronic, magnetic, optical and mechanical properties. These important qualities start from electronics where the moiré superlattice led to Hofstadter's butterfly[4-6], superconductivity[7], correlated insulating states[8] and many others[9-13]. Later on, ferromagnetism[14] and moiré excitons[15-18] were reported in vdW structures at small magic twisting angles. In nanophotonics and polaritonic nano-optics where the research involves highly confined light-matter waves called polaritons[19, 20], the moiré superlattice was shown to reflect plasmon polaritons in graphene[21-24] and produce nano-light photonic crystals[21]. These previous breakthroughs in configuring vdW materials remark the peculiarities in the formed moiré superlattice as a special case of vdW configuration at small specific twisting angles δ (typically δ < 5º). To further exploit the advances of vdW systems, it is worth exploring the configurability via twisting, stacking etc. beyond the small specific δ range. Yet, for polaritons in vdW systems, without evident moiré superlattice, they remain rigid at a broad range of twisting angles (δ > 5º). This lack of configuration effect stems from the spreading nature of polaritons in most vdW materials: once excited, the polaritons propagate toward all directions in the basal plane. The insufficient electromagnetic anisotropy of the spreading propagation manner weakens the twisting configuration by rotating the components.

In this work, we report on the configuration of polaritons in twisted α-phase molybdenum trioxide (α-MoO$_3$) slabs. Different from other spreading polaritons, polaritons in α-MoO$_3$[25, 26] are extremely anisotropic: they propagate only along certain directions in the basal plane. Using real-space infrared (IR) nano-imaging of the propagating polaritons, we demonstrated various wavefront geometry and polariton topological transitions by twisting stacked α-MoO$_3$ slabs with an angle δ from 0° to 90° (Fig. 1(a)). The polariton twisting configuration originates from the electromagnetic hybridization between directional polaritons in the top and bottom α-MoO$_3$ slab where the hybridization strongly depends on the twisting angle δ. Since a variety of polariton wavefront geometry like open hyperbola and parenthesis, closed oval and squircle as well as their deformation including compression and stretching can all be obtained simply by varying the twisting angle δ, twisted α-MoO$_3$ in this work holds promises to offer tailorable IR nano-light for various nano-optical functionalities.

The IR nano-imaging of polaritons in twisted α-MoO$_3$ were performed using scattering-type scanning near-field optical microscopy (s-SNOM). The s-SNOM is based on a tapping-mode atomic force microscope (AFM) which simultaneously yields the topography and nano-IR image of the scanned area (Fig. 1(a)). In the experiment, the AFM tip is illuminated with an IR laser (brown solid arrow) at the frequency ω = 1 / λ$_0$ (λ$_0$ is the wavelength of IR light in free space), the topography and back scattered near-field signal (brown dashed arrow) are recorded during the scan. The experimental observable near-field amplitude $S(\omega)$ and phase $\Phi(\omega)$ typically possess a spatial resolution ~ 10 nm, close to the radius of the AFM tip, and therefore can map nano-polaritons in the real-space[19, 20, 27]. The polaritons are typically imaged as fringes – standing wave interference oscillations – in the s-SNOM experiments.

At the IR frequency $\omega = 915$ cm$^{-1}$, a representative s-SNOM amplitude $S(\omega)$ image of the single α-MoO$_3$ slab is shown in Fig. 1(b). In the experiment, we fabricated an Au disk (thickness 100 nm, diameter 1 μm) on top of the α-MoO$_3$ slab. Two types of polariton fringes – the $S(\omega)$ oscillations – can be observed: "X" shape hyperbolic fringes around the Au disk (red disk) and linear fringes close to the slab edge (white dashed line). The linear fringes originate from the interference between polaritons launched by the s-SNOM tip and those reflected at the crystal edges. These fringes are always parallel to the crystal edge[28-30] regardless of the polariton wavefront geometry. The "X" shape fringes around the Au disk are interference patterns between polaritons launched by the Au disk and the IR illumination (brown solid arrow in Fig. 1(a))[29, 31]. For highly confined polaritons in vdW materials[29], these fringes superposition with the wavefront of the launched polaritons and therefore are mainly discussed in this work. In contrast to spreading concentric fringes from other vdW materials, "X" shape fringes observed in α-MoO$_3$ indicate the extreme anisotropy of the directional polaritons. This polariton directionality stems from the hyperbolic response ($\varepsilon_i \varepsilon_j < 0$, where $\varepsilon_i$ and $\varepsilon_j$ are permittivity along different directions, i and j denote crystal axis [100], [001] or [010]) inside Reststrahlen bands of α-MoO$_3$[25, 26]: Band 1 at $\omega = 545$ to $851$ cm$^{-1}$ for [001] phonon ($\varepsilon_{[100]} > 0$, $\varepsilon_{[001]} < 0$ and $\varepsilon_{[010]} > 0$), Band 2 at $\omega = 820$ to $972$ cm$^{-1}$ for [100] phonon ($\varepsilon_{[100]} < 0$, $\varepsilon_{[001]} > 0$ and $\varepsilon_{[010]} > 0$), and Band

3 at ω = 958 to 1010 cm$^{-1}$ for [010] phonon ($\varepsilon_{[100]} > 0$, $\varepsilon_{[010]} > 0$ and $\varepsilon_{[010]} < 0$). Specifically, the directional "X" shape fringes (Fig. 1(b)) are directly attributed to the basal plane hyperbolic response ($\varepsilon_{[100]}\varepsilon_{[001]} < 0$) inside the Reststrahlen band 2: the isofrequency dispersion for the α-MoO₃ slab is an open hyperbola (Fig. 1(c), green curve) in the basal plane and polaritons are propagating along fixed directions (green arrows in Fig. 1(c)) with an angle $\theta = arc\,tan\sqrt{|\varepsilon_{[100]}/\varepsilon_{[010]}|}$ to the [010] direction [26, 32].

The directional polaritons in the basal plane suggest the configurability by twisting stacked α-MoO₃ slabs inside the Reststrahlen band 2. In Figures 1(d)–(h), we provided s-SNOM images of polaritons in twisted α-MoO₃ at various twisting angle δ. The s-SNOM images are oriented with all [001] axis of the bottom α-MoO₃ slab aligned in the vertical direction (see the coordinate axis in Figure 1). At δ = 0°, the polaritons exhibit the "X" shape fringes (Fig. 1(d)) that correspond to the hyperbolic wavefront, very similar to that of the single slab α-MoO₃ (Fig. 1(b)). At δ = 20° (Fig. 1(e)), the hyperbolic wavefront can still be observed but is titled from that with δ = 0° and the single slab. The polariton wavefront at δ = 63° is different: the fringes can only be observed along a tilted line across the Au disk (Fig. 1(f)), the wavefront is therefore in a parenthesis geometry. With this parenthesis wavefront, polaritons are confined along the tilted line, the propagation along other directions are prohibited. This parenthesis wavefront is also revealed in polariton fringes launched by a defect (red * in Fig. 1(f)) where similar parenthesis fringes along the same tilted direction can be observed. At δ = 81° (Fig. 1(g)), the polariton wavefront becomes oval as fringes observed around the Au dot. Finally, at δ = 90° (Fig. 1(h)), the wavefront geometry appears as a squircle, rounded square polariton fringes are observed. In Fig.

1(h), the squircle fringes are less visible on the lower left side of the Au disk, due to the shadowing effect of the AFM cantilever[29].

s-SNOM images in Figure 1 reveal various wavefront geometry and topological transitions of the polaritons in twisted α-MoO₃. As the twisting angle δ increases from 0º to 90º, the polariton wavefront exhibit a hyperbola (Fig. 1(d), δ = 0º), tilted hyperbola (Fig. 1(e), δ = 20º), parenthesis (Fig. 1(f), δ = 63º), oval (Fig. 1(g), δ = 81º) and squircle (Fig. 1(h), δ = 90º). Note that at δ ~ 63º, the polariton wavefront changes from an open hyperbola-like geometry (Figures 1(d)-(e)) to a closed ellipse-like geometry (diamond, squircle in Figures 1(g)-(h)), corresponding to the topological transitions[33, 34]. This continuous polariton twisting configuration in α-MoO₃ over the broad range δ = 0º to 90º is in stark contrast to plasmons in twisted graphene and graphene/hBN heterostructures where plasmon polaritons are affected at specific and small twisting angles (δ < 5º) by the moiré superlattice[21-23].

The observed polariton twisting configuration is supported both by our finite element method (FEM) simulation and electromagnetic wave theory in Figure 2. With the input of α-MoO₃ slab thickness and IR frequency ω from our experiments and fitted α-MoO₃ permittivity based on ref. [25, 26, 35], we simulated the real-space electromagnetic field $E_z$ and modelled momentum space (*k*-space) isofrequency dispersion in twisted α-MoO₃. In the real-space simulation (see details in supplementary information Section 1), a vertical dipole was placed at the center (Figures 2(a)-(f), top) and above the twisted α-MoO₃ as the polariton launcher. The real-space simulation reproduces the topological transition of polaritons observed in our experiments (Figure 1): the wavefront exhibit the hyperbola at the single slab and δ = 0º (Fig. 2 (a), (b), top), tilted hyperbola at δ = 20º (Fig. 2(c), top), parenthesis at δ = 63º (Fig. 2(d), top), oval at δ = 81º (Fig. 2(e), top) and squircle at δ = 90º (Fig. 2(f), top). Our electromagnetic theory (see details in Supplementary Information Section 2) of the isofrequency dispersion in twisted α-MoO$_3$ was developed from previous theoretical works on black phosphorus[36, 37] and graphene metasurface[33, 38]. The modeled *k*-space isofrequency curves (Figures 2(a)-(f), bottom) are in excellent agreement with the real-space images produced in our experiments (Figure 1) and FEM simulations (Figures 2(a)-(f), top).

We attribute the twisting configuration reported in Figures 1-2 to electromagnetic interaction between extremely anisotropic polaritons in the stacked top and bottom α-MoO$_3$ slab. Specifically, the twisting configuration requires 1) extreme basal plane anisotropy of the polariton and 2) sufficient overlap of polariton field from stacked slabs. In Figure 3, we provide experimental data and calculation analysis to demonstrate these criteria for the twisting configuration. At a representative IR frequency ω = 980 cm$^{-1}$ inside the Reststrahlen Band 3, the *k*-space polariton geometry is an ellipse (red curve, Fig. 1(c)) and polaritons propagate with an elliptical wavefront in the real-space (Fig. 3(a)). Compared with directional polaritons inside the Band 1 (Fig. 1(b)), polaritons at ω = 980 cm$^{-1}$ lack sufficient anisotropy such that they spread towards all directions (Fig. 3(a)). The twisting configuration is ineffective at ω = 980 cm$^{-1}$: polaritons from twisted α-MoO$_3$ with δ = 20º (Fig. 3(b)), 63º (Fig. 3(c)) and 90º (Fig. 3(d)) all exhibit elliptical wavefront similar to that in the single slab (Fig. 3(a)). These results are in stark contrast with the s-SNOM image at ω = 900 cm$^{-1}$ (Fig. 3(f)) where the squircle and a series of other wavefront (Figure 1) can be configured by twisting α-MoO$_3$ slabs supporting the directional polaritons.

In addition to the extreme basal plane anisotropy, sufficient field overlap between polaritons in stacked vdW structures is also required to build adequate electromagnetic interaction to configure polaritons. Since polaritons exponentially decay out-side-of the α-MoO$_3$ slab, this criterion corresponds to a moderate decay of the polariton field. In Fig. 3(g), we calculated the distribution of electromagnetic field $|E_z|$ away from the α-MoO$_3$ slab (see Supplementary Information Section 2 for details) at three representative IR frequencies: ω = 900 cm$^{-1}$ (green) and 935 cm$^{-1}$ (blue) inside the Band 2 and ω = 980 cm$^{-1}$ (red) inside the Band 3. Away from the α-MoO$_3$ slab, the polariton field $|E_z|$ decays quickly at ω = 935 and 980 cm$^{-1}$ while moderately at ω = 900 cm$^{-1}$: the field decay length where $|E_z|$ becomes $1/e$ of that at $z = 0$ is 14, 92 and 243 nm for ω = 980, 935 and 900 cm$^{-1}$, respectively. Therefore, in stacked α-MoO$_3$, the polaritons field overlap between the top and bottom α-MoO$_3$ is sufficient at ω = 900 cm$^{-1}$ while insufficient at ω = 935 and 980 cm$^{-1}$. The sufficient polariton field overlap leads to adequate electromagnetic interaction and thus the twisting configuration at ω = 900 cm$^{-1}$. Our s-SNOM data in Figures 3(d)-(f) support this theoretical expectation. Evident twisting configuration was observed at ω = 900 cm$^{-1}$ (Fig. 3(f) and Figure 1) while the lack of twisting configuration was observed at ω = 935 and 980 cm$^{-1}$. Note that at ω = 935 cm$^{-1}$ (Fig. 3(e)), the polaritons in the top and bottom α-MoO$_3$ slab fulfill the other criterion to be basal plane anisotropic (hyperbolic wavefront), yet these hyperbolic modes barely interact with each other due to their insufficient field overlap (blue curve in Fig. 3(g)). Adequate electromagnetic interaction as the criterion of polariton twisting configuration is further confirmed by our simulation results of separated α-MoO$_3$ slabs in Supplementary Section 3.

Combined experimental and theoretical results in Figures 1-3 report the configuration of nano-polaritons in twisted α-MoO$_3$. The polariton wavefront can be altered by controlling the twisting angle δ between stacked α-MoO$_3$ slabs in a broad range of 0° to 90°, thus demonstrating the extended vdW configurability of the nanoscale electromagnetic energy beyond Morié engineering with small twisting angles [4-13, 21, 22]. The observed twisting configuration of polaritons is attributed to electromagnetic interaction between anisotropic polariton fields where the interaction is highly dependent on the twisting angle. Note that by configuring the wavefront via twisting, the tuning of polariton propagation direction and wavelength surpasses previous modulation efforts through refractive index engineering[39, 40] or vdW heterostructuring[41-43]. Future work may be guided towards exploring the dynamics and reconfigurability of twisted polariton nano-light via nano-mechanical manipulation[2] or vdW photonic hybrids[1, 41-44]. Various polariton wavefront and topologies demonstrated in this work suggest the opportunity to offer polariton nano-light with tailored propagating properties and photonic density of states[45, 46] for on-demand nano-optical functionalities that can benefit light emission[46], quantum optics[47, 48] and exotic transitions[49] etc. The twisting configuration of hyperbolic polaritons in α-MoO$_3$ also provides a prototype for the exploration of configuring other anisotropic physical properties in vdW materials via twisting, stacking and heterostructuring.

**Acknowledgments:** We acknowledge helpful discussions with X. Jiang, J. Lin and T. Low. Work in the PJH group was supported through AFOSR grant FA9550-16-1-0382 (fabrication), and the Gordon and Betty Moore Foundations EPiQS Initiative through Grant GBMF4541 to PJH. This work made use of the Materials Research Science and Engineering Center Shared Experimental Facilities supported by the National Science Foundation (NSF) (Grant No. DMR-0819762). The work at Zhejiang University was sponsored by the National Natural Science Foundation of China (NNSFC) under Grants No. 61625502, No.11961141010, and No. 61975176, the Top-Notch Young Talents Program of China, and the Fundamental Research Funds for the Central Universities.

**Reference**

[1] A.K. Geim, I.V. Grigorieva, Van der Waals heterostructures, Nature, 499 (2013) 419.

[2] R. Ribeiro-Palau, C. Zhang, K. Watanabe, T. Taniguchi, J. Hone, C.R. Dean, Twistable electronics with dynamically rotatable heterostructures, Science, 361 (2018) 690-693.

[3] K.S. Novoselov, A. Mishchenko, A. Carvalho, A.H. Castro Neto, 2D materials and van der Waals heterostructures, Science, 353 (2016) aac9439.

[4] L.A. Ponomarenko, R.V. Gorbachev, G.L. Yu, D.C. Elias, R. Jalil, A.A. Patel, A. Mishchenko, A.S. Mayorov, C.R. Woods, J.R. Wallbank, M. Mucha-Kruczynski, B.A. Piot, M. Potemski, I.V. Grigorieva, K.S. Novoselov, F. Guinea, V.I. Fal'ko, A.K. Geim, Cloning of Dirac fermions in graphene superlattices, Nature, 497 (2013) 594.

[5] C.R. Dean, L. Wang, P. Maher, C. Forsythe, F. Ghahari, Y. Gao, J. Katoch, M. Ishigami, P.

Moon, M. Koshino, T. Taniguchi, K. Watanabe, K.L. Shepard, J. Hone, P. Kim, Hofstadter's butterfly and the fractal quantum Hall effect in moiré superlattices, Nature, 497 (2013) 598.

[6] B. Hunt, J.D. Sanchez-Yamagishi, A.F. Young, M. Yankowitz, B.J. LeRoy, K. Watanabe, T. Taniguchi, P. Moon, M. Koshino, P. Jarillo-Herrero, R.C. Ashoori, Massive Dirac fermions and Hofstadter butterfly in a van der Waals heterostructure, Science, 340 (2013) 1427-1430.

[7] Y. Cao, V. Fatemi, S. Fang, K. Watanabe, T. Taniguchi, E. Kaxiras, P. Jarillo-Herrero, Unconventional superconductivity in magic-angle graphene superlattices, Nature, 556 (2018) 43.

[8] Y. Cao, V. Fatemi, A. Demir, S. Fang, S.L. Tomarken, J.Y. Luo, J.D. Sanchez-Yamagishi, K. Watanabe, T. Taniguchi, E. Kaxiras, R.C. Ashoori, P. Jarillo-Herrero, Correlated insulator behaviour at half-filling in magic-angle graphene superlattices, Nature, 556 (2018) 80.

[9] L. Wang, Y. Gao, B. Wen, Z. Han, T. Taniguchi, K. Watanabe, M. Koshino, J. Hone, C.R. Dean, Evidence for a fractional fractal quantum Hall effect in graphene superlattices, Science, 350 (2015) 1231-1234.

[10] R. Krishna Kumar, X. Chen, G.H. Auton, A. Mishchenko, D.A. Bandurin, S.V. Morozov, Y. Cao, E. Khestanova, M. Ben Shalom, A.V. Kretinin, K.S. Novoselov, L. Eaves, I.V. Grigorieva, L.A. Ponomarenko, V.I. Fal'ko, A.K. Geim, High-temperature quantum oscillations caused by recurring Bloch states in graphene superlattices, Science, 357 (2017) 181-184.

[11] M. Lee, J.R. Wallbank, P. Gallagher, K. Watanabe, T. Taniguchi, V.I. Fal'ko, D. Goldhaber-Gordon, Ballistic miniband conduction in a graphene superlattice, Science, 353 (2016) 1526-1529.

[12] Z. Shi, C. Jin, W. Yang, L. Ju, J. Horng, X. Lu, H.A. Bechtel, M.C. Martin, D. Fu, J. Wu, K.

Watanabe, T. Taniguchi, Y. Zhang, X. Bai, E. Wang, G. Zhang, F. Wang, Gate-dependent pseudospin mixing in graphene/boron nitride moiré superlattices, Nature Physics, 10 (2014) 743.

[13] E.M. Spanton, A.A. Zibrov, H. Zhou, T. Taniguchi, K. Watanabe, M.P. Zaletel, A.F. Young, Observation of fractional Chern insulators in a van der Waals heterostructure, Science, 360 (2018) 62-66.

[14] A.L. Sharpe, E.J. Fox, A.W. Barnard, J. Finney, K. Watanabe, T. Taniguchi, M.A. Kastner, D. Goldhaber-Gordon, Emergent ferromagnetism near three-quarters filling in twisted bilayer graphene, Science, 365 (2019) 605-608.

[15] K. Tran, G. Moody, F. Wu, X. Lu, J. Choi, K. Kim, A. Rai, D.A. Sanchez, J. Quan, A. Singh, J. Embley, A. Zepeda, M. Campbell, T. Autry, T. Taniguchi, K. Watanabe, N. Lu, S.K. Banerjee, K.L. Silverman, S. Kim, E. Tutuc, L. Yang, A.H. MacDonald, X. Li, Evidence for moiré excitons in van der Waals heterostructures, Nature, 567 (2019) 71-75.

[16] E.M. Alexeev, D.A. Ruiz-Tijerina, M. Danovich, M.J. Hamer, D.J. Terry, P.K. Nayak, S. Ahn, S. Pak, J. Lee, J.I. Sohn, M.R. Molas, M. Koperski, K. Watanabe, T. Taniguchi, K.S. Novoselov, R.V. Gorbachev, H.S. Shin, V.I. Fal'ko, A.I. Tartakovskii, Resonantly hybridized excitons in moiré superlattices in van der Waals heterostructures, Nature, 567 (2019) 81-86.

[17] C. Jin, E.C. Regan, A. Yan, M. Iqbal Bakti Utama, D. Wang, S. Zhao, Y. Qin, S. Yang, Z. Zheng, S. Shi, K. Watanabe, T. Taniguchi, S. Tongay, A. Zettl, F. Wang, Observation of moiré excitons in WSe2/WS2 heterostructure superlattices, Nature, 567 (2019) 76-80.

[18] K.L. Seyler, P. Rivera, H. Yu, N.P. Wilson, E.L. Ray, D.G. Mandrus, J. Yan, W. Yao, X. Xu, Signatures of moiré-trapped valley excitons in MoSe2/WSe2 heterobilayers, Nature, 567 (2019)

66-70.

[19] D.N. Basov, M.M. Fogler, F.J. García de Abajo, Polaritons in van der Waals materials, Science, 354 (2016) aag1992.

[20] T. Low, A. Chaves, J.D. Caldwell, A. Kumar, N.X. Fang, P. Avouris, T.F. Heinz, F. Guinea, L. Martin-Moreno, F. Koppens, Polaritons in layered two-dimensional materials, Nature Materials, 16 (2016) 182.

[21] G.X. Ni, H. Wang, J.S. Wu, Z. Fei, M.D. Goldflam, F. Keilmann, B. Özyilmaz, A.H. Castro Neto, X.M. Xie, M.M. Fogler, D.N. Basov, Plasmons in graphene moiré superlattices, Nature Materials, 14 (2015) 1217.

[22] S.S. Sunku, G.X. Ni, B.Y. Jiang, H. Yoo, A. Sternbach, A.S. McLeod, T. Stauber, L. Xiong, T. Taniguchi, K. Watanabe, P. Kim, M.M. Fogler, D.N. Basov, Photonic crystals for nano-light in moire graphene superlattices, Science, 362 (2018) 1153-1156.

[23] L. Jiang, Z. Shi, B. Zeng, S. Wang, J.-H. Kang, T. Joshi, C. Jin, L. Ju, J. Kim, T. Lyu, Y.-R. Shen, M. Crommie, H.-J. Gao, F. Wang, Soliton-dependent plasmon reflection at bilayer graphene domain walls, Nature Materials, 15 (2016) 840.

[24] F. Hu, S.R. Das, Y. Luan, T.F. Chung, Y.P. Chen, Z. Fei, Real-Space Imaging of the Tailored Plasmons in Twisted Bilayer Graphene, Physical Review Letters, 119 (2017) 247402.

[25] W. Ma, P. Alonso-González, S. Li, A.Y. Nikitin, J. Yuan, J. Martín-Sánchez, J. Taboada-Gutiérrez, I. Amenabar, P. Li, S. Vélez, C. Tollan, Z. Dai, Y. Zhang, S. Sriram, K. Kalantar-Zadeh, S.-T. Lee, R. Hillenbrand, Q. Bao, In-plane anisotropic and ultra-low-loss polaritons in a natural van der Waals crystal, Nature, 562 (2018) 557-562.

[26] Z. Zheng, N. Xu, S.L. Oscurato, M. Tamagnone, F. Sun, Y. Jiang, Y. Ke, J. Chen, W. Huang, W.L. Wilson, A. Ambrosio, S. Deng, H. Chen, A mid-infrared biaxial hyperbolic van der Waals crystal, Science Advances, 5 (2019) eaav8690.

[27] T.G. Folland, L. Nordin, D. Wasserman, J.D. Caldwell, Probing polaritons in the mid- to far-infrared, Journal of Applied Physics, 125 (2019) 191102.

[28] S. Dai, Z. Fei, Q. Ma, A.S. Rodin, M. Wagner, A.S. McLeod, M.K. Liu, W. Gannett, W. Regan, K. Watanabe, T. Taniguchi, M. Thiemens, G. Dominguez, A.H.C. Neto, A. Zettl, F. Keilmann, P. Jarillo-Herrero, M.M. Fogler, D.N. Basov, Tunable Phonon Polaritons in Atomically Thin van der Waals Crystals of Boron Nitride, Science, 343 (2014) 1125-1129.

[29] S. Dai, Q. Ma, Y. Yang, J. Rosenfeld, M.D. Goldflam, A. McLeod, Z. Sun, T.I. Andersen, Z. Fei, M. Liu, Y. Shao, K. Watanabe, T. Taniguchi, M. Thiemens, F. Keilmann, P. Jarillo-Herrero, M.M. Fogler, D.N. Basov, Efficiency of Launching Highly Confined Polaritons by Infrared Light Incident on a Hyperbolic Material, Nano Letters, 17 (2017) 5285-5290.

[30] G.X. Ni, A.S. McLeod, Z. Sun, L. Wang, L. Xiong, K.W. Post, S.S. Sunku, B.Y. Jiang, J. Hone, C.R. Dean, M.M. Fogler, D.N. Basov, Fundamental limits to graphene plasmonics, Nature, 557 (2018) 530-533.

[31] P. Alonso-González, A.Y. Nikitin, F. Golmar, A. Centeno, A. Pesquera, S. Vélez, J. Chen, G. Navickaite, F. Koppens, A. Zurutuza, F. Casanova, L.E. Hueso, R. Hillenbrand, Controlling graphene plasmons with resonant metal antennas and spatial conductivity patterns, Science, 344 (2014) 1369-1373.

[32] S. Dai, Q. Ma, T. Andersen, A.S. McLeod, Z. Fei, M.K. Liu, M. Wagner, K. Watanabe, T.

Taniguchi, M. Thiemens, F. Keilmann, P. Jarillo-Herrero, M.M. Fogler, D.N. Basov, Subdiffractional focusing and guiding of polaritonic rays in a natural hyperbolic material, Nature Communications, 6 (2015) 6963.

[33] J.S. Gomez-Diaz, M. Tymchenko, A. Alù, Hyperbolic Plasmons and Topological Transitions Over Uniaxial Metasurfaces, Physical Review Letters, 114 (2015) 233901.

[34] H.N.S. Krishnamoorthy, Z. Jacob, E. Narimanov, I. Kretzschmar, V.M. Menon, Topological Transitions in Metamaterials, Science, 336 (2012) 205.

[35] G. Álvarez-Pérez, T.G. Folland, I. Errea, J. Taboada-Gutiérrez, J. Duan, J. Martín-Sánchez, A.I.F. Tresguerres-Mata, J.R. Matson, A. Bylinkin, M. He, W. Ma, Q. Bao, J.D. Caldwell, A.Y. Nikitin, P. Alonso-González, Infrared permittivity of the biaxial van der Waals semiconductor α-$MoO_3$ from near- and far-field correlative studies, arXiv e-prints, 2019, pp. arXiv:1912.06267.

[36] M. Renuka, X. Lin, Z. Wang, L. Shen, B. Zheng, H. Wang, H. Chen, Dispersion engineering of hyperbolic plasmons in bilayer 2D materials, Opt. Lett., 43 (2018) 5737-5740.

[37] A. Nemilentsau, T. Low, G. Hanson, Anisotropic 2D Materials for Tunable Hyperbolic Plasmonics, Physical Review Letters, 116 (2016) 066804.

[38] G. Hu, A. Krasnok, Y. Mazor, C.-W. Qiu, A. Alù, Moiré Hyperbolic Metasurfaces, arXiv e-prints, 2020, pp. arXiv:2001.03304.

[39] K. Chaudhary, M. Tamagnone, M. Rezaee, D.K. Bediako, A. Ambrosio, P. Kim, F. Capasso, Engineering phonon polaritons in van der Waals heterostructures to enhance in-plane optical anisotropy, Science Advances, 5 (2019) eaau7171.

[40] A. Fali, S.T. White, T.G. Folland, M. He, N.A. Aghamiri, S. Liu, J.H. Edgar, J.D. Caldwell, R.F. Haglund, Y. Abate, Refractive Index-Based Control of Hyperbolic Phonon-Polariton Propagation, Nano Letters, 19 (2019) 7725-7734.

[41] V.W. Brar, M.S. Jang, M. Sherrott, S. Kim, J.J. Lopez, L.B. Kim, M. Choi, H. Atwater, Hybrid Surface-Phonon-Plasmon Polariton Modes in Graphene/Monolayer h-BN Heterostructures, Nano Letters, 14 (2014) 3876-3880.

[42] S. Dai, Q. Ma, M.K. Liu, T. Andersen, Z. Fei, M.D. Goldflam, M. Wagner, K. Watanabe, T. Taniguchi, M. Thiemens, F. Keilmann, G.C.A.M. Janssen, S.E. Zhu, P. Jarillo-Herrero, M.M. Fogler, D.N. Basov, Graphene on hexagonal boron nitride as a tunable hyperbolic metamaterial, Nature Nanotechnology, 10 (2015) 682.

[43] A. Woessner, M.B. Lundeberg, Y. Gao, A. Principi, P. Alonso-González, M. Carrega, K. Watanabe, T. Taniguchi, G. Vignale, M. Polini, J. Hone, R. Hillenbrand, F.H.L. Koppens, Highly confined low-loss plasmons in graphene–boron nitride heterostructures, Nature Materials, 14 (2015) 421-425.

[44] J.D. Caldwell, I. Vurgaftman, J.G. Tischler, O.J. Glembocki, J.C. Owrutsky, T.L. Reinecke, Atomic-scale photonic hybrids for mid-infrared and terahertz nanophotonics, Nature Nanotechnology, 11 (2016) 9-15.

[45] Z. Jacob, J.Y. Kim, G.V. Naik, A. Boltasseva, E.E. Narimanov, V.M. Shalaev, Engineering photonic density of states using metamaterials, Applied Physics B, 100 (2010) 215-218.

[46] T. Galfsky, J. Gu, E.E. Narimanov, V.M. Menon, Photonic hypercrystals for control of light–matter interactions, Proceedings of the National Academy of Sciences, 114 (2017) 5125.

[47] T.B. Hoang, G.M. Akselrod, M.H. Mikkelsen, Ultrafast Room-Temperature Single Photon Emission from Quantum Dots Coupled to Plasmonic Nanocavities, Nano Letters, 16 (2016) 270-275.

[48] S.I. Bogdanov, M.Y. Shalaginov, A.S. Lagutchev, C.-C. Chiang, D. Shah, A.S. Baburin, I.A. Ryzhikov, I.A. Rodionov, A.V. Kildishev, A. Boltasseva, V.M. Shalaev, Ultrabright Room-Temperature Sub-Nanosecond Emission from Single Nitrogen-Vacancy Centers Coupled to Nanopatch Antennas, Nano Letters, 18 (2018) 4837-4844.

[49] N. Rivera, I. Kaminer, B. Zhen, J.D. Joannopoulos, M. Soljačić, Shrinking light to allow forbidden transitions on the atomic scale, Science, 353 (2016) 263.

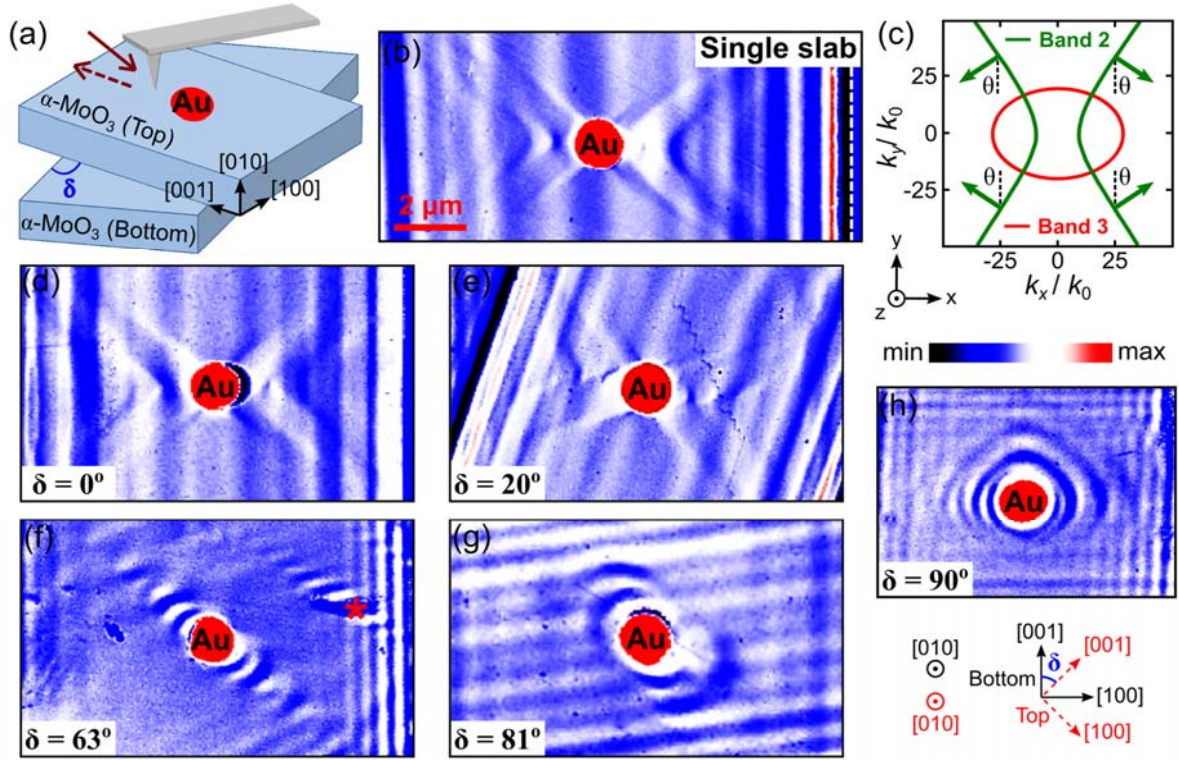

FIG. 1. Real-space IR nano-images reveal the twisting configuration for nano-polaritons. (a) Experiment schematic. Solid and dashed brown arrows denote the incident and back-scattered IR light. (b) s-SNOM amplitude image of single slab α-MoO₃. (c) Isofrequency curve of hyperbolic polaritons in Reststrahlen Band 2 and Band 3. Green arrows denote the momentum of polaritons in Band 2. (d) – (h) s-SNOM amplitude images of twisted α-MoO₃ with δ = 0°, 20°, 63°, 81° and 90°. IR frequency in b): 915 cm$^{-1}$, d): 915 cm$^{-1}$, e): 910 cm$^{-1}$, f): 920 cm$^{-1}$, g): 915 cm$^{-1}$, h): 905 cm$^{-1}$. Scale bar: 2 μm. Note that at the fixed frequency ω = 910 and 915 cm$^{-1}$, the s-SNOM images (Supplementary Section 4) of twisted α-MoO₃ show the same twisting configuration and topological transitions of polaritons as those in Figure 1.

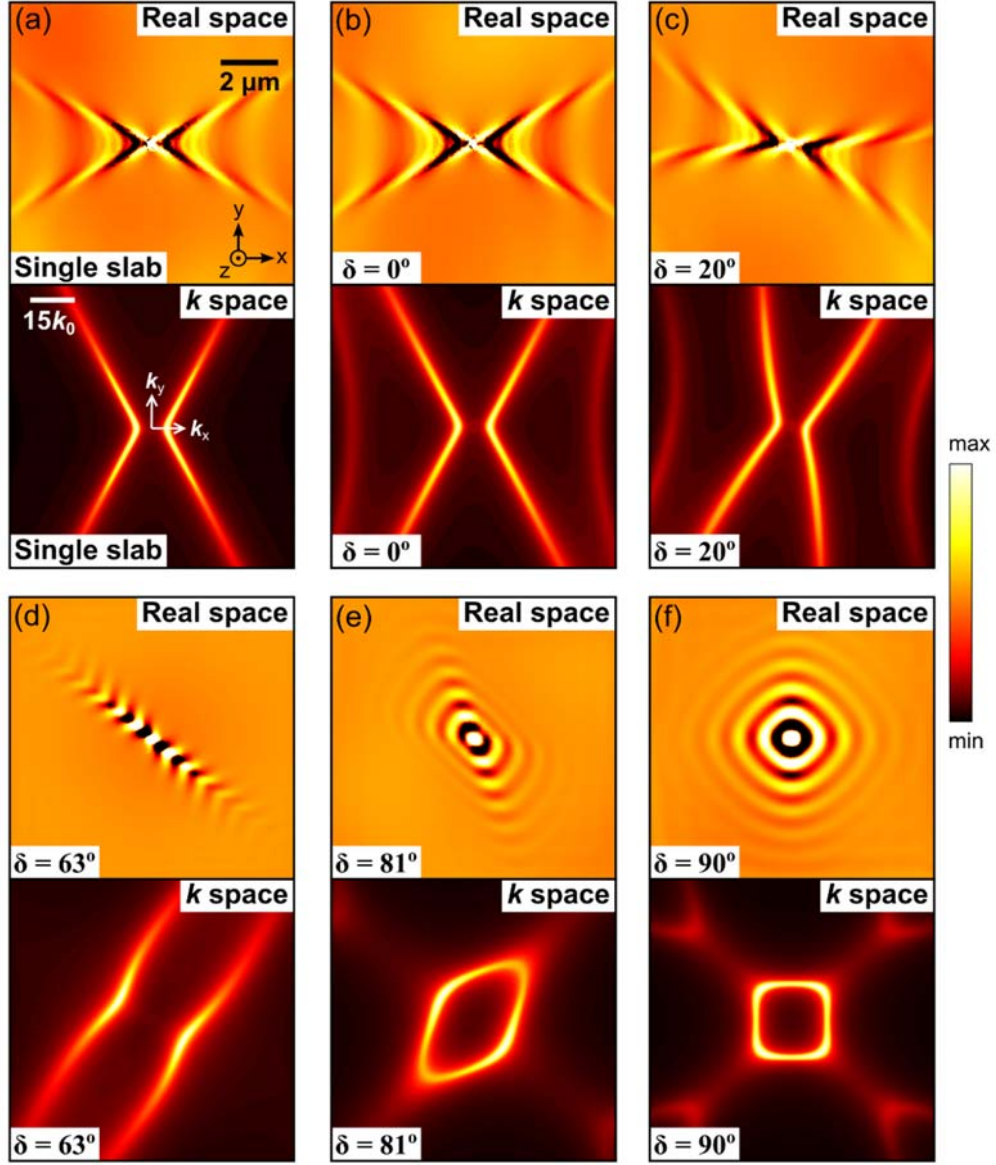

FIG. 2. Finite Element Method (FEM) real-space simulation and electromagnetic theory of momentum-space (*k*-space) isofrequency dispersion of twisted polaritons. (a) – (f) top, real-space FEM simulation of polaritons in twisted α-MoO$_3$ where a dipole was placed above the center of the image to launch the polaritons. Scale bar: 2 μm. (a) – (f) bottom, electromagnetic theory of corresponding isofrequency dispersions. Scale bar: 15 $k_0$, $k_0$ is the momentum of IR photon $k_0 = 2\pi / \lambda_0$.

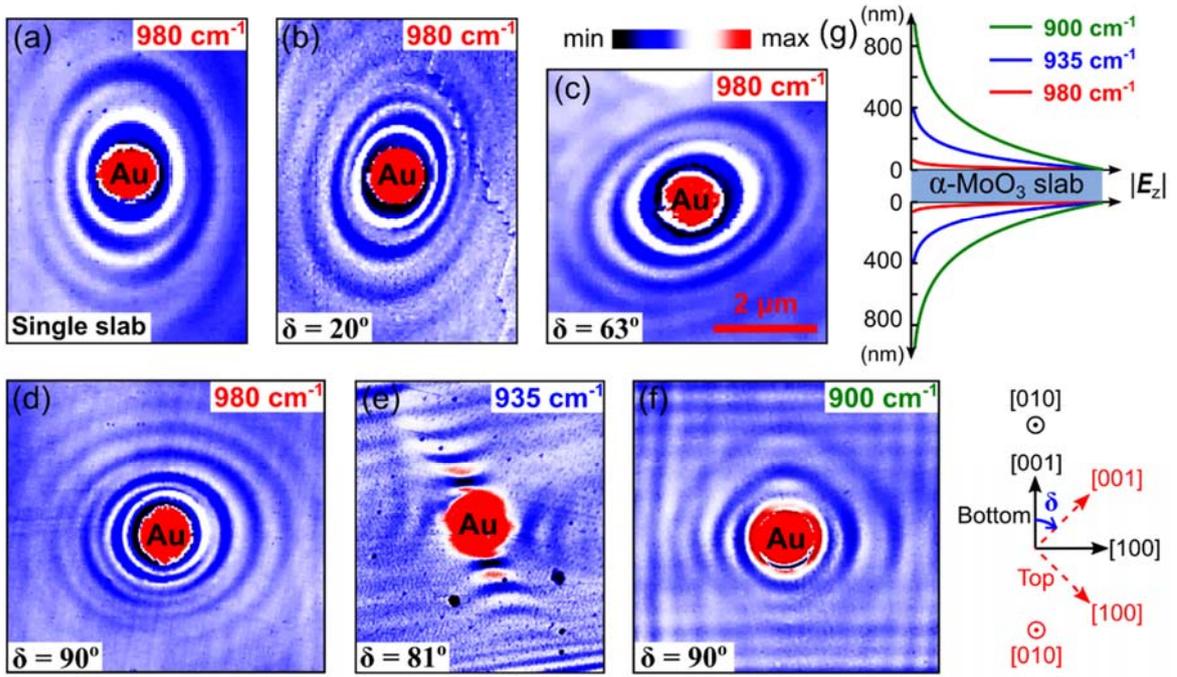

FIG. 3. Electromagnetic interaction as the origin of twisting configuration in polaritons. (a) – (d) s-SNOM amplitude images of single slab α-MoO3 (a) and twisted α-MoO3 with δ = 20⁰, 63⁰ and 90⁰ at IR frequency 980 cm⁻¹. (e) s-SNOM amplitude images of twisted α-MoO3 with δ = 81⁰ at IR frequency 935 cm⁻¹. (f) s-SNOM amplitude images of twisted α-MoO3 with δ = 90⁰ at IR frequency 900 cm⁻¹. Scale bar: 2 μm. (g), theoretical electromagnetic field distribution $|E_z|$ along the [100] direction away from the α-MoO3 slab. α-MoO3 thickness: 100 nm.